# RETURN TO VENUS OF THE JAPANESE VENUS CLIMATE ORBITER AKATSUKI

**Masato Nakamura**
Institute of Space and Astronautical Science, Japan, nakamura.masato@jaxa.jp

Yasuhiro Kawakatsu[*], Chikako Hirose[*], Takeshi Imamura[*], Nobuaki Ishii[*], Takumi Abe[*], Atsushi Yamazaki[*], Manabu Yamada[†], Kazunori Ogohara[*], Kazunori Uemizu[*], Tetsuya Fukuhara[‡], Shoko Ohtsuki[§], Takehiko Satoh[*], Makoto Suzuki[*], Munetaka Ueno[*], Junichi Nakatsuka[*], Naomoto Iwagami[**], Makoto Taguchi[††], Shigeto Watanabe[‡], Yukihiro Takahashi[‡], George L. Hashimoto[‡‡], and Hiroki Yamamoto[*]

Japanese Venus Climate Orbiter/AKATSUKI was proposed in 2001 with strong support by international Venus science community and approved as an ISAS (The Institute of Space and Astronautical Science) mission soon after the proposal. The mission life we expected was more than two Earth years in Venus orbit. AKATSUKI was successfully launched at 06:58:22JST on May 21, 2010, by H-IIA F17. After the separation from H-IIA, the telemetry from AKATSUKI was normally detected by DSN Goldstone station (10:00JST) and the solar cell paddles' deployment was confirmed. After a successful cruise, the malfunction happened on the propulsion system during the Venus orbit insertion (VOI) on Dec 7, 2010. The engine shut down before the planned reduction in speed to achieve. The spacecraft did not enter the Venus orbit, but entered an orbit around the Sun with a period of 203 days. Most of the fuel still had remained, but the orbital maneuvering engine was found to be broken and unusable. However, we have found an alternate way of achieving orbit by using only the reaction control system (RSC). We had adopted the alternate way for orbital maneuver and three minor maneuvers in Nov 2011 were successfully done so that AKATSUKI would meet Venus in 2015. We are considering several scenarios for VOI using only RCS.

## I. INTRODUCTION

Venus is our nearest neighbor and has a size very similar to the Earth's; however, previous spacecraft missions discovered an extremely dense (~92 bar) and dry $CO_2$ atmosphere with $H_2SO_4$-$H_2O$ clouds floating at high altitudes, and exotic volcanic features covering the whole planet. The abundant gaseous $CO_2$ brings about a high atmospheric temperature (~740 K) near the surface via greenhouse effect. The atmospheric circulation is also much different from the Earth's. The mechanisms that sustain such conditions are unclear and considered to be the keys to understanding the origin of the terrestrial environment.

[*] Institute of Space and Astronautical Science, Japan, surname.givenname@jaxa.jp
[†] Planetary Exploration Research Center, Chiba Institute of Technology, Japan, manabu@perc.it-chiba.ac.jp
[‡] Hokkaido University, Japan,
  tetsuyaf@ep.sci.hokudai.ac.jp,
  shw@ep.sci.hokudai.ac.jp,
  yukihiro@mail.sci.hokudai.ac.jp,
[§] Senshu University, Japan, oh@isc.senshu-u.ac.jp
[**] University of Tokyo, Japan, iwagami@eps.s.u-tokyo.ac.jp
[††] Rikkyo University, Japan, taguchi@rikkyo.ac.jp
[‡‡] Okayama University, Japan, george@gfd-dennou.org

Japanese Venus Climate Orbiter/AKATSUKI was proposed in 2001 to address the mysterious Venusian atmosphere with strong support by international Venus science community and approved as an ISAS mission soon after the proposal [1, 2]. PLANET-C was the given project code name in ISAS. AKATSUKI and ESA's Venus Express have been expected to complement each other in Venus climate study. Various coordinated observations using the two spacecraft have been planned. Also participating scientists from the U.S. have been selected. The mission life we expected was more than two Earth years in Venus orbit.

AKATSUKI was successfully launched at 06:58:22JST on May 21, 2010, by H-IIA F17. After the separation from H-IIA, the telemetry from AKATSUKI was normally detected by DSN Goldstone station and the solar cell paddles' deployment was confirmed. AKATSUKI was put into the three-axis stabilized mode in the initial operation from Uchinoura station and the critical operation was finished at 20:00JST on the same day.

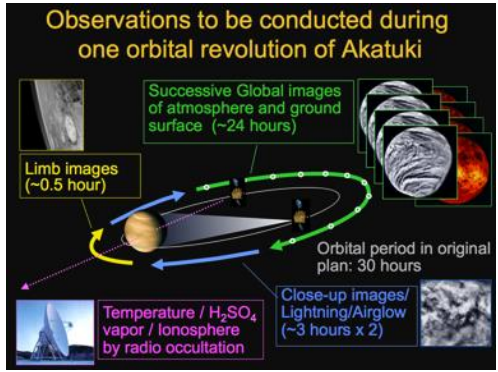

Fig. I: Original orbital plan of AKATSUKI and the observation schedule.

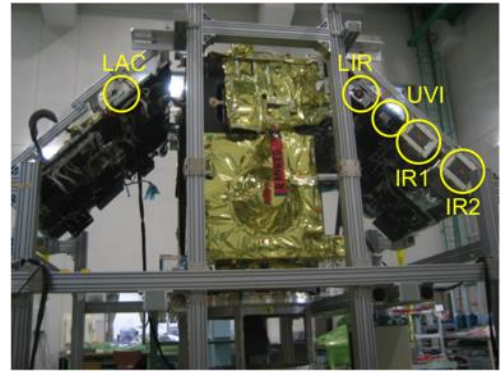

Fig. II: Spacecraft and mission instruments.

## II. MISSION OF AKATSUKI

AKATSUKI aims to elucidate the mechanism of the mysterious atmospheric circulation of Venus, with secondary targets being the exploration of the ground surface. The exploration of the Venusian meteorology is given a high priority not only for understanding the climate of Venus but also for the general understanding of planetary fluid dynamics. The systematic imaging sequence of AKATSUKI is advantageous for detecting meteorological phenomena with various temporal and spatial scales. The long elliptical orbit is suitable for obtaining cloud-tracked wind vectors over a wide area continuously from high altitudes (Fig. I). With such wind data, the characterizations of the meridional circulation, mid-latitude jets, and various wave activities are anticipated.

We have five photometric sensors as mission instruments, which are 1μm-Infrared camera (IR1), 2μm-Infrared camera (IR2), Ultra-Violet Imager (UVI), Longwave Infrared camera (LIR), and Lightning and Airglow Camera (LAC) (Fig. II).

IR1 is designed to image the dayside of Venus at 0.90 μm wavelength and the nightside at 0.90, 0.97, and 1.01 μm wavelengths, which are located in the atmospheric windows [3, 4]. The measurements at 0.90 and 1.01 μm will yield information about the surface material [5, 6].

IR2 utilizes the atmospheric windows at wavelengths of 1.73, 2.26, and 2.32 μm: the first two are nearly absorption free, while the last one contains a CO absorption band. At these wavelengths, IR2 is most sensitive to infrared radiation originating from altitudes 35-50 km. To track cloud motions, a series of 2.26-μm images is exclusively used. As the inhomogeneity of the Venusian cloud layer is thought to occur predominantly at altitudes 50-55 km [7], the IR2 observations should yield wind maps in this region. IR2 employs two additional wavelengths, 2.02 μm (a prominent $CO_2$ absorption band) and an astronomical H-band centered at 1.65 μm. At 2.02 μm, we expect to detect variations of cloud-top altitude as intensity variations of reflected sunlight.

UVI is designed to measure ultraviolet radiation scattered from cloud tops at ~65 km altitude in two bands centered at 283 and 365 nm wavelengths. The Venusian atmosphere shows broad absorption of solar radiation between 200 and 500 nm. The absorption in the range between 200 and 320 nm is explained by $SO_2$ at the cloud top, while the absorption above 320 nm should be due to another absorber that is not identified yet [8].

LIR detects thermal emission from the cloud top in a rather wide wavelength region 8-12 μm to map the cloud-top temperature [9, 10]. Unlike other imagers onboard AKATSUKI, LIR is able to take images of both dayside and nightside with equal quality and accuracy. The cloud-top temperature map will reflect the cloud height distribution, whose detailed structure is unknown except in the northern high latitudes observed by the infrared radiometer onboard Pioneer Venus [11], as well as the atmospheric temperature distribution.

LAC is a high-speed imaging sensor, which measures lightning flashes and airglow emissions on the nightside disk of Venus when AKATSUKI is located within the umbra (shadow region) of Venus [12]. One of the major goals of LAC is to settle controversy on the occurrence of lightning in the Venusian atmosphere. We will also obtain information on the global circulation in the lower thermosphere by continuous observations of large-scale structures in the $O_2$ Herzberg II (552.5 nm) night airglow, whose production is a consequence of the recombination of atomic oxygen in downwelling.

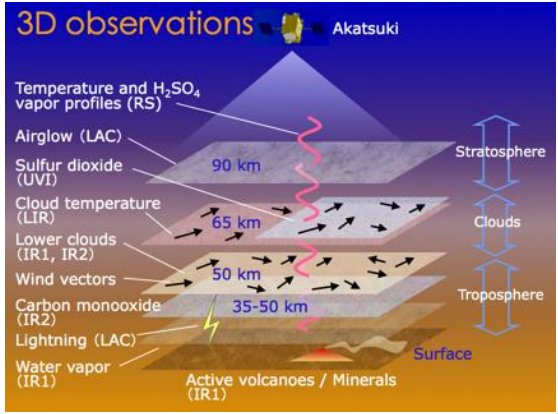

Fig. III: Three-dimensional observation of the atmosphere by mission instruments.

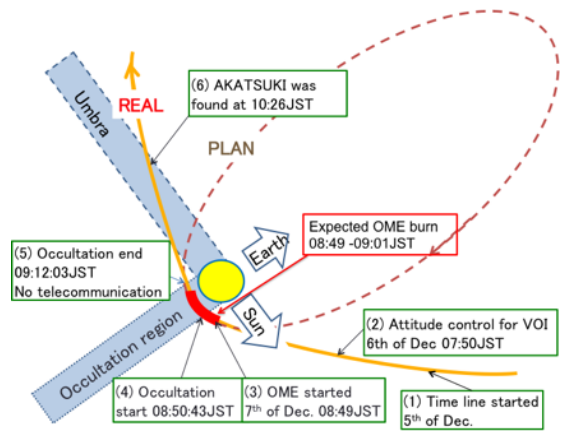

Fig. IV: VOI trial on Dec 7, 2012.

In addition to the photometric observations above, radio occultation experiments obtain high-resolution vertical profiles of the temperature, the sulfuric acid density, and the ionospheric electron density [13]. In this observation mode, when the spacecraft is occulted by Venus as seen from the Earth, radio waves transmitted by the spacecraft traverse the Venusian atmosphere, reach the tracking station, and are recorded for offline analysis. For this experiment the spacecraft is equipped with an ultra-stable oscillator.

Figure III shows the schematic image of three-dimensional observation of the atmosphere by the combination of these instruments.

## III. THE FAILURE OF VENUS ORBIT INSERTION

The Venus orbit insertion (VOI) of AKATSUKI had been scheduled for Dec 7, 2010; however, the spacecraft did not enter the orbit around Venus due to the malfunction of the propulsion system.

The time sequence before and after the VOI is as follows (Fig. IV).
(1) Time line commands on the spacecraft started on Dec 5.
(2) Attitude control maneuver was completed at 07:50JST on Dec 6.
(3) Orbital maneuvering engine (OME) started at 08:49JST on Dec 7.
(4) The spacecraft occultation (seen from the Earth) started at 08:50:43JST on the same day and no telemetry was available until the end of the expected occultation end at 09:12:03JST.
(5) After the occultation had ended, we tried to get the telemetry from the spacecraft but failed.
(6) The search of AKATSUKI had been intensively done by ISAS and NASA's Deep Space Network managed by JPL, and finally it was found at 10:26JST at an unexpected position, which was not on the planned orbit around Venus. The spacecraft had automatically switched to the safe hold mode. It means something unexpected had happened on the spacecraft during the occultation.

Later investigation based on the analysis from the telemetry data recorded in the occultation period revealed that the OME was stopped at 08:51:38JST, which is about 10 minutes before the schedule (the nominal OME stop timing was scheduled at 09:01:00JST). At 158 seconds after OME start, the attitude of the spacecraft suddenly inclined due to the relatively large disturbance. Then, the attitude control system stopped its three-axis control and switched to the safe mode in which one axis perpendicular to the solar power paddles was directed toward the sun and the spacecraft body was slowly rotated about the axis.

The investigation by JAXA and the Ministry of Education, Culture, Sports, Science and Technology of Japan[*] over half a year after the VOI failure revealed that the check valve between the helium tank and the fuel tank (Fig. V) was blocked by salt for-mation during the journey from the Earth to Venus. The valve motion was inspected in the huge number of ground tests in the mixed vapor of the fuel and the oxidizer. We found that the mixed gases under the low temperature produced the salt formation on the valve spindle. The check valve was blocked by the salt formation and did not work. Then, the OME became oxidizer-rich and fuel-poor condition, which leads to the abnormal combustion in the engine with high temperature. Finally, the temperature exceeded the upper limit and the throat of the engine was broken. The thrust change and a quite large transversal force caused by the engine damage reproduced in the ground tests were consistent to the change of the probe's acceleration observed on the orbit.

---

[*] http://www.mext.go.jp/b_menu/shingi/uchuu/reports/1317561.htm

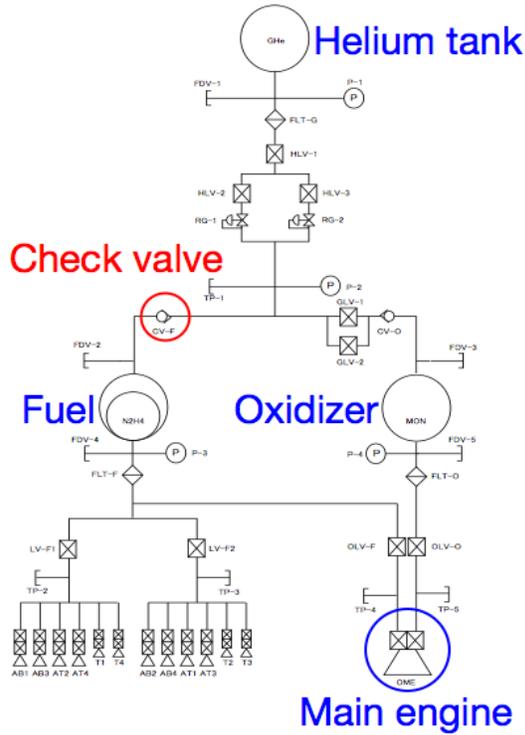

Fig. V: Propulsion system on AKATSUKI.

## IV. PRESENT STATUS OF AKATSUKI

As described in the previous section the Venus orbit insertion scheduled for Dec 2010 has failed due to a malfunction of the main thruster. Now, AKATSUKI is orbiting the Sun. All the subsystems, except the main thruster, are normal. We are carefully monitoring the long-term change of their temperatures.

From the result of the ground tests and the telemetry data that show the acceleration just before the stop of the combustion was about 70% of the nominal value, we considered OME's nozzle skirt was broken but the chamber throat might remain usable in a blow-down mode. In this case, a reduction of the ignition spike is required to prevent a further damage to the chamber. We preformed further combustion tests on the ground for about 200 times and found that we can reduce the ignition spike by delaying the injection timing of the oxidizer and controlling the temperature of the injector and valves in a certain range. After a rehearsal for controlling the temperature of the spacecraft, we conducted a test of the thruster on the orbit in Sep 2011. However, the obtained thrust value was about 10% of our expectations. We have speculated that the remaining chamber throat had been damaged more than we had considered and it was broken by the reduced ignition spike of this test.

Considering the above result, we decided not to use the main thruster anymore and to use the attitude control thrusters (or the reaction control system, RCS) for further orbit maneuver. RCS does not require oxidizer and we disposed the oxidizer of 65 kg in Oct 2011. Reducing the spacecraft weight enables us to perform further orbital maneuver easier.

We decided to try the second VOI operation by AKATSUKI in 2015, which is the timing that the spacecraft will come close to Venus on the present orbit around the Sun. We performed orbit control maneuvers with Delta-V of about 243.8 m/s in total on Nov 1, 10, and 21, 2011. With this operation, the spacecraft will make a rendezvous with Venus in Nov 2015.

Our main concern is the high temperature condition of the spacecraft at the perihelion (0.6 AU) where the heat flux into the spacecraft per unit area is 38% higher (3655 W/m$^2$) than the expected highest heat flux (2649 W/m$^2$) on the orbit around Venus (Fig. VI).

Several scientific observations were conducted during the extended cruise phase. Images of Venus were obtained by UVI, IR1, and LIR from a distance of ~$6\times10^5$ km on Dec 9, 2010, which is two days after the failure of VOI (Fig. VII). The mid-infrared images taken by LIR revealed the existence of zonal belts and patchy structures in the cloud temperature

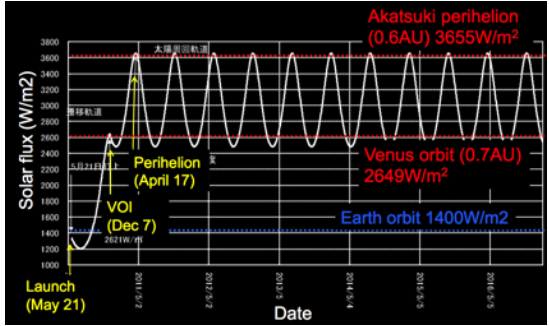

Fig. VI: Expected heat flux into the spacecraft during the cruise phase.

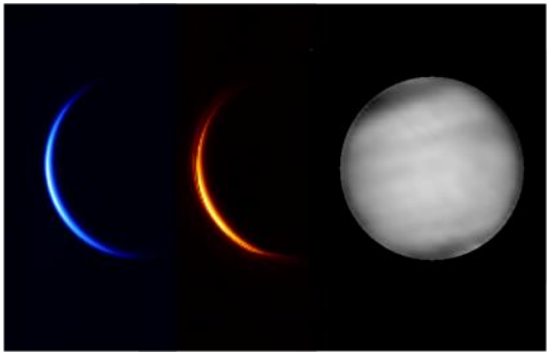

Fig. VII: Images of Venus nightside taken by UVI (wavelength: 283 nm; left), IR1 (0.9 μm; center), and LIR (10 μm; right) on Dec 9, 2010.

distribution [14]; the findings shed light on the mysteries of Venus' cloud dynamics.

Photometric observations of Venus were conducted also in Mar and May 2011 using IR1, IR2, UVI, and LIR. The intensity variations during this period reflect the change in the observation geometry and the temporal variation of the Venusian atmosphere. The dependence of the near-infrared reflectivity on the solar scattering angle was successfully determined using the IR1 and IR2 data and was used to constrain the cloud structure. The result suggests the existence of anomalously large particles in the upper cloud region, implying an unknown temporal variation in the cloud system [15]. In addition to the scattering angle dependence, quasi-periodic variations with periods of ~4 days were clearly seen in the ultraviolet reflectivity determined by UVI. This variation is attributed to the planetary-scale inhomogeneities of the ultraviolet absorbers embedded in the super-rotational atmosphere (4-day rotation). The amplitudes of the variations at 283 and 365 nm wavelengths and the phase difference between them constrain the photochemistry and dynamics in the cloud layer. The mid-infrared brightness taken by LIR also shows distinct variations with time scales of several days, which are under study.

The solar conjunction in Jun 2011 provided a unique opportunity to explore the solar corona by radio occultation technique. The observations covered heliocentric distances of 1.5-20 solar radii and revealed radial variations of the solar wind velocity, the activity of compressional waves, and the turbulence density spectrum.

V. ORBIT INSERTION PLAN IN 2015

The characteristic of Venus is that its axial tilt is 177.4 degrees. This means that Venus rotates clock-wise (retrograde) viewed from above the north pole of the Sun, whereas the Earth rotates counter-clockwise (direct). As observation requirements of the Venus circular orbit, two things should be considered: one is that the orbital plane should be close to the Venus equator and the other is that the rotation direction should be the same direction as the super-rotation of the Venusian atmosphere.

In order to achieve the above possible orbits with retrograde direction in the close Venus equator, several possible strategies were investigated. However, almost of them were proved not to be applicable, because the periapsis altitude decreased and the orbit finally collided with the planetary surface due to the solar gravitational disturbance. Currently, two possible orbits turned out to satisfy the previous requirements [16]. One of them uses the gravity brake method, and the other utilizes the Hohmann transfer method; the two orbits are described in the following two paragraphs.

The first candidate orbit is designed based on gravity brake method, whose key idea is the maximum use of the solar perturbation. Blue arrows in Fig. VIII indicate the direction of acceleration by the solar perturbation in the Sun-Venus fixed rotating frame, where the Sun always positions at left of Venus. When a probe rotates retrograde, the solar perturbation accelerates the probe in the first and third quadrants (upper right and lower left areas) and it decelerates the probe in the second and fourth quadrants (upper left and lower right areas). The designed trajectory is shown by the colored curve both in the Sun-Venus fixed frame (Fig. VIII) and in the inertial frame (Fig. IX). We insert the spacecraft initially to the direct orbit in the deceleration area (the second quadrant in Fig. VIII; dark blue curve labeled by "2. direct"), but we choose the apoapsis altitude of 1 million km and make it fly up to just inside the Hill radius of Venus. By this, we make the most of the solar perturbation of deceleration and make the spacecraft rotate inversely from direct to retrograde during this period (dark blue one labeled by "3. retrograde"). Moreover, when it flies back near Venus again, the trajectory moves from the first to the fourth quadrant where the perturbation acts as from deceleration to acceleration (red and light blue ones labeled by "4-6"). This makes it possible to keep the periapsis altitude in the Venus circular orbit, which was impossible in the usual orbit insertion. The spacecraft performs a couple of maneuvers afterward to reach an elliptical orbit with major axis of hundreds of thousands kilometers and minor axis of a few thousand kilometers.

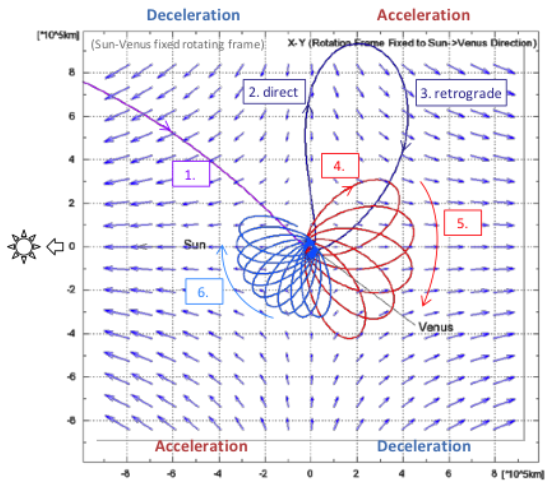

Fig. VIII: The trajectory for the gravity break method in the Sun-Venus fixed rotating frame.

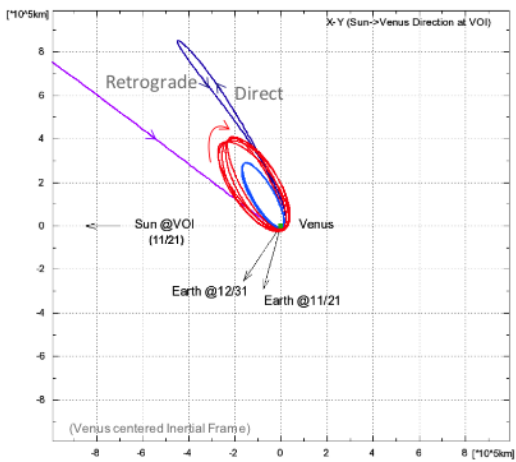

Fig. IX: The trajectory for the gravity break method in the inertial frame.

The second possible orbit is designed based on Hohmann transfer method, in which the orbit of AKATSUKI contacts that of Venus internally. For this method, AKATSUKI has to perform a deceleration maneuver of approximately 80 m/s before the Venus encounter in 2015 to change the approach angle largely in order that the circular orbit begins in the acceleration area (the first quadrant in Fig. VIII) to prevent from decreasing the periapsis. Because of the fuel consumption of 80 m/s before the orbit insertion, the final apoapsis altitude of this orbit will be a hundred thousand kilometers higher than that of the orbit with the gravity break method..

For the safe Venus orbit insertion, we are presently investigating the sensitivity to trajectory errors in the two methods. Furthermore, since AKATSUKI flies in the same orbital plane with the Venus, it is also difficult to prevent the spacecraft from shadow (umbra) in either approach. After confirming the safety and trade-off of the two methods, one recovery plan will be fixed in 2013.

## VI. CONCLUDING SUMMARY

AKATSUKI aims at understanding the mechanism of atmospheric circulation of Venus, with additional targets being the exploration of the ground surface. This paper has described brief introduction of AKATSUKI mission and its mission instruments, the mission status after the launch in May 2010, as well as the recovery plan of the mission by inserting the spacecraft into the orbit around Venus in 2015. Even in the severe condition that the orbital maneuvering engine is unable to use, we are planning the Venus orbit insertion (VOI) again only with the reaction control system. We should achieve success even in such a condition with the rest of the fuel. The project team is carefully operating the spacecraft to keep the spacecraft in the best condition until the day of the next VOI.

[2] Nakamura, M., T. Imamura, N. Ishii, T. Abe, T. Satoh, M. Suzuki, M. Ueno, A. Yamazaki, N. Iwagami, S. Watanabe, M. Taguchi, T. Fukuhara, Y. Takahashi, M. Yamada, N. Hoshino, S. Ohtsuki, K. Uemizu, G. L. Hashimoto, M. Takagi, Y. Matsuda, K. Ogohara, N. Sato, Y. Kasaba, T. Kouyama, N. Hirata, R. Nakamura, Y. Yamamoto, N. Okada, T. Horinouchi, M. Yamamoto, and Y. Hayashi, 2011. Overview of Venus orbiter, Akatsuki. *Earth Planets Space* **63**, 443-457.

[3] Taylor, F. W., D. Crisp, and B. Bézard, 1997. Near-infrared souding of the lower atmosphere of Venus. In: Bougher, S. W., D. M. Hunten, and R. J. Phillips (Eds.), *Venus II*. University of Arizona Press.

[4] Iwagami. N., S. Takagi, S. Ohtsuki, M. Ueno, K. Uemizu, T. Satoh, T. Sakanoi, and G. L. Hashimoto, 2011. Science requirements and description of the 1 μm camera onboard the Akatsuki Venus Orbiter. *Earth Planets Space* **63**, 487-492.

[5] Baines, K. H., G. Bellucci, J. P. Bibring, R. H. Brown, B. J. Buratti, E. Bussoletti, F. Capaccioni, P. Cerroni, R. N. Clark, A. Coradini, D. P. Cruikshank, P. Drossart, V. Formisano, R. Jaumann, Y. Langevin, D. L. Matson, T. B. McCord, V. Mennella, R. M. Nelson, P. D. Nicholson, B. Sicardy, C. Sotin, G. B. Hansent, J. J. Aiello, and S. Amici, 2000. Detection of sub-micron radiation from the surface of Venus by Cassini/VIMS. *Icarus* **148**, 307-311.

[6] Hashimoto, G. L., S. Sugita, 2003. On observing the compositional variability of the surface of Venus using nightside near-infrared thermal radiation. *J. Geophys. Res.* **108**, 5109, doi:10.1029/2003JE002082.

[7] Belton, M. J. S., P. J. Gierasch, M. D. Smith, P. Helfenstein, P. J. Schinder, J. B. Pollack, K. A. Rages, A. P. Ingersoll, K. P. Klaasen, J. Veverka, C. D. Anger, M. H. Carr, C. R. Chapman, M. E. Davies, F. P. Fanale, R. Greeley, R. Greenberg, J. W. Head III, D. Morrison, G. Neukum, and C. B. Pilcher, 1991. Imaging from Galileo of the Venus Cloud Deck. *Science* **253**, 1531-1536.

[8] Esposito, L. W., J.-L. Bertaux, V. Krasnopolsky, V. I. Moroz, and L. V. Zasova, 1997. Chemistry of lower atmosphere and clouds. In: Bougher, S. W., D. M. Hunten, and R. J. Phillips (Eds.), *Venus II*. University of Arizona Press.

[9] Taguchi, M., T. Fukuhara, T. Imamura, M. Nakamura, N. Iwagami, M. Ueno, M. Suzuki, G. L. Hashimoto, and K. Mitsuyama, 2007. Longwave infrared camera onboard the Venus Climate Orbiter. *Adv. Space Res.*, **40**, 861-868.

[10] Fukuhara, T., M. Taguchi, T. Imamura, M. Nakamura, M. Ueno, M. Suzuki, N. Iwagami, M. Sato, K. Mitsuyama, G. L. Hashimoto, R. Ohshima, T. Kouyama, H. Ando, and M. Futaguchi, 2011. LIR: Longwave Infrared Camera onboard the Venus orbiter Akatsuki. *Earth Planets Space,* **63**, 1009-1018.

[11] Taylor, F. W., R. Beer, M. T. Chahine, D. J. Diner, L. S. Elson, R. D. Haskins, D. J. McCleese, J. V. Martonchik, P. E. Reichley, S. P. Bradley, J. Delderfield, J. T. Schofield, C. B. Farmer, L. Froidevaux, J. Leung, M. T. Coffey, and J. C. Gille, 1980. Structure and meteorology of the middle atmosphere of Venus: infrared remote sounding from the Pioneer Orbiter. *J. Geophys. Res.* **85**, 7963-8006.

[12] Takahashi, Y., J. Yoshida, Y. Yair, T. Imamura, and M. Nakamura, 2008. Lightning detection by LAC onboard the Japanese Venus climate orbiter, Planet-C. *Space Sci. Rev.* **137**, 317-334.

[13] Imamura, T., T. Toda, A. Tomiki, D. Hirahara, T. Hayashiyama, N. Mochizuki, Z. Yamamoto, T. Abe, T. Iwata, H. Noda, Y. Futaana, H. Ando, B. Häusler, M. Pätzold, and A. Nabatov, 2011. Radio occultation experiment of the Venus atmosphere and ionosphere with the Venus orbiter Akatsuki. *Earth Planets Space* **63**, 493-501.

[14] Taguchi, M., T. Fukuhara, M. Futaguchi, M. Sato, T. Imamura, K. Mitsuyama, M. Nakamura, M. Ueno, M. Suzuki, N. Iwagami, and G. L. Hashimoto, 2012. Characteristic features in Venus' nightside cloud-top temperature obtained by Akatsuki/LIR. *Icarus* **219**, 502-504.

[15] Satoh, T., S. Ohtsuki, N. Iwagami, M. Ueno, K. Uemizu, M. Suzuki, G. L. Hashimoto, T. Sakanoi, Y. Kasaba, R. Nakamura, T. Imamura, M. Nakamura, T. Fukuhara, A. Yamazaki, and M. Yamada, Venus' clouds as inferred from the phase curves acquired by IR1 and IR2 on board Akatsuki. submitted to *Icarus*.

[16] Hirose, C., N. Ishii, Y. Kawakatsu, C. Ukai, and H. Terada, 2012. The trajectory control strategies for Akatsuki re-insertion into the Venus orbit. *Proceedings of 23rd International Symposium on Space Flight Dynamics*.
IAC-12-A3.5.3 − Page 7 of 7